\documentclass[pra,twocolumn,superscriptaddress,10pt]{revtex4-1}
\usepackage{amsmath,amssymb,bm}
\usepackage{xcolor}

\newtheorem{lemma}{Lemma}
\newtheorem{theorem}{Theorem}
\newenvironment{proof}
{\par\noindent\textit{Proof.}\ }
{\hfill$\square$\par}

\begin{document}

\title{
Kempe-Invariant Ordering of Three-Qubit Pairwise Entanglement under Local Depolarization
}
\author{Wei Song}\email{wsong1@mail.ustc.edu.cn}
\affiliation{School of Physics and Materials Engineering, Hefei Normal University,
Hefei 230601, China}

\author{Xiao-Lan Zong}\email{zxl@hfnu.edu.cn}
\affiliation{School of Physics and Materials Engineering, Hefei Normal University,
Hefei 230601, China}

\author{Ming Yang}
\affiliation{School of Physics, Anhui University, Hefei, 230601, China}
\affiliation{Leibniz International Joint Research Center of Materials Sciences of Anhui Province, Anhui University, Hefei 230601, China}

\date{\today}

\begin{abstract}
Pure three-qubit states with the same pairwise concurrences and three-tangle can still have different Kempe invariants. We show that the Kempe invariant orders the entanglement dynamics of all three pairs under identical, independent local depolarizing channels, with smaller values giving greater robustness. The same ordering also holds for the sum of squared pairwise entanglements of formation and the sudden-death thresholds of pairwise entanglement
\end{abstract}

\maketitle

\section{Introduction}
\label{sec:introduction}

For a pure three-qubit state, the three pairwise concurrences \cite{Wootters1998} and the
three-tangle \cite{CKW2000} provide four entanglement coordinates, but do not generally
determine the state up to local unitaries(LU). At generic interior points, fixing these coordinates
leaves one continuous degree of freedom, which can be parametrized by the Kempe invariant \cite{Kempe1999,Sudbery2001,Acin2000,Osterloh2010,Song2026Compatibility}

The physical role of LU invariants has been explored in studies of both correlations and
entanglement dynamics. Cheng and Hall related anisotropic invariants of three-qubit states
to pairwise correlations, Bell nonlocality, and quantum steering
\cite{ChengHall2017}; these invariance properties were subsequently tested
experimentally \cite{Zhu2019}. The robustness of multipartite entanglement
under local decoherence, including independent depolarization, has also been
studied \cite{SimonKempe2002,Borras2009}. Li and Zhao found different
robustness even for LU-equivalent states under amplitude damping, dephasing,
and bit-flip noise \cite{LiZhao2018}, where the channels can distinguish
local orientations. For local depolarization, Zhang, Jiang, and Liang showed
that the initial decay of one-versus-rest negativity depends on LU-invariant
information beyond the usual tangle coordinates \cite{Zhang2013}.

Here we ask whether the Kempe invariant orders the full concurrence trajectories of all three pairs for pure
three-qubit states with fixed initial pairwise concurrences and three-tangle. We consider identical, independent
local depolarizing channels, whose LU covariance ensures that these trajectories are independent of the choice
of local basis. We prove that the initial Kempe invariant simultaneously orders all three pairwise concurrence
trajectories, with smaller values corresponding to greater entanglement robustness. Thus, the Kempe invariant
orders the robustness of pairwise entanglement among states with the same four initial entanglement coordinates.

\section{Invariant fibers of pure three-qubit states}
\label{sec:fibers}
We use
\begin{equation}
\bm E=(C_{AB},C_{AC},C_{BC},t),
\qquad
t\equiv\tau_3,
\label{eq:Evector}
\end{equation}
to specify the four initial entanglement coordinates considered here. At a
generic interior point, fixing $\bm E$ leaves one continuous invariant
coordinate \cite{Osterloh2010,Song2026Compatibility}. We call the corresponding
allowed interval of $\kappa$ the \emph{Kempe fiber}. Its endpoints are given in
Sec.~\ref{sec:consequences}; on nongeneric strata, the interval
may collapse to a single point.

A point on this fiber need not correspond to a unique LU orbit.
In particular, a state and its complex conjugate can have the same five
real invariants while being LU inequivalent \cite{Acin2001}.
These states nevertheless have identical concurrence dynamics under
local depolarization, since the channel commutes with complex
conjugation and $C(\rho^*)=C(\rho)$. More generally, at any fixed
noise strength, the expressions in Appendices~\ref{app:reduction}
and \ref{app:polynomial} determine the spin-flip spectrum of each
reduced pair, and hence its concurrence, from $(\bm E,\kappa)$ alone.

For the pair $AB$, we use the Bloch representation
\begin{equation}
\begin{split}
\rho_{AB}
=
\frac14\bigg[
&I\otimes I
+\bm r_A\cdot\bm\sigma\otimes I
+I\otimes\bm r_B\cdot\bm\sigma
\\
&+\sum_{m,n=1}^{3}
T^{AB}_{mn}\sigma_m\otimes\sigma_n
\bigg].
\end{split}
\label{eq:blochAB}
\end{equation}
The other two pairs are written in the same way. We define
\begin{equation}
R_i=|\bm r_i|^2,
\qquad
K_{ij}=\|T^{ij}\|_F^2,
\qquad
U_{ij}=\bm r_i^T T^{ij}\bm r_j.
\label{eq:basicinvariants}
\end{equation}

For $\{i,j,k\}=\{A,B,C\}$, the CKW relations give
\begin{equation}
R_i
=
1-C_{ij}^2-C_{ik}^2-t.
\label{eq:Ri}
\end{equation}

For a pure three-qubit state, $\rho_{ij}$ and $\rho_k$ have the same
nonzero spectrum, so that
\begin{equation}
\operatorname{Tr}\rho_{ij}^2
=
\operatorname{Tr}\rho_k^2.
\end{equation}

Using the Bloch representation, we have
\begin{equation}
K_{ij}
=
1+2R_k-R_i-R_j.
\label{eq:Kij}
\end{equation}

Thus, fixing $\bm E$ fixes the lengths of the three local Bloch vectors
and the Frobenius norms of the three pair-correlation tensors.

For any reduced pair $ij$, the Kempe invariant can be written as
\begin{equation}
\kappa
=
3\operatorname{Tr}[(\rho_i\otimes\rho_j)\rho_{ij}]
-\operatorname{Tr}\rho_i^3
-\operatorname{Tr}\rho_j^3.
\label{eq:kempe}
\end{equation}

In the Pauli representation,
\begin{equation}
U_{AB}=U_{AC}=U_{BC}\equiv U
=
\frac{4\kappa-1}{3}.
\label{eq:Ukempe}
\end{equation}

Thus the three reduced pairs share the same invariant $U$.
At fixed $\bm E$, we use this common variable to compare their
concurrence dynamics under local depolarization.

\section{Local depolarization and the Wootters root}
\label{sec:channel}
We let each qubit evolve under the same local depolarizing channel,
\begin{equation}
\Lambda_\eta(\rho)
=
\eta\rho+(1-\eta)\frac{I}{2},
\qquad
0\le\eta\le1.
\label{eq:depol}
\end{equation}

Since the local channel on the third qubit is trace preserving,
the reduced state of the pair $ij$ evolves as
\begin{equation}
\rho_{ij}(\eta)
=
(\Lambda_\eta\otimes\Lambda_\eta)(\rho_{ij}).
\label{eq:pairchannel}
\end{equation}

Setting $s=\eta^2$, we have
\begin{equation}
\begin{split}
\rho_{ij,s}
=
\frac14\bigg[
&I\otimes I
+\sqrt{s}\,\bm r_i\cdot\bm\sigma\otimes I
+\sqrt{s}\,I\otimes\bm r_j\cdot\bm\sigma\\
&+s\sum_{m,n}
T^{ij}_{mn}\sigma_m\otimes\sigma_n
\bigg].
\end{split}
\label{eq:rhosgeneral}
\end{equation}

At fixed $\bm E$ and noise strength, the one- and two-qubit purities
are independent of $\kappa$. To determine the $\kappa$ dependence
of concurrence, we examine the spin-flip spectrum of each reduced pair.

For a given pair, define
\begin{equation}
\widetilde{\rho}_{ij,s}
=
(\sigma_y\otimes\sigma_y)
\rho_{ij,s}^{*}
(\sigma_y\otimes\sigma_y),
\qquad
M_{ij,s}
=
\rho_{ij,s}\widetilde{\rho}_{ij,s}.
\label{eq:spinflipM}
\end{equation}

Denote the eigenvalues of $M_{ij,s}$ by $\mu_\alpha$ and set
\begin{equation}
r_\alpha=\sqrt{\mu_\alpha},
\qquad
r_1\ge r_2\ge r_3\ge r_4\ge0.
\end{equation}

The concurrence is then
\begin{equation}
C_{ij,\eta}
=
\max\{0,r_1-r_2-r_3-r_4\}.
\label{eq:woottersgeneral}
\end{equation}

The quartic equation for the spin-flip eigenvalues can be solved
analytically \cite{AzumaBan2010}. To study how the concurrence
depends on $\kappa$, we introduce the auxiliary polynomial

\begin{equation}
\Pi_{ij}(q)
=
\prod_{\alpha=1}^{4}
\left[
q-
\left(
2r_\alpha-\sum_{\beta=1}^{4}r_\beta
\right)^2
\right].
\label{eq:Pidef}
\end{equation}
The coefficients depend only on symmetric combinations of the
eigenvalues of $M_{ij,s}$; the explicit expressions are listed in
Appendix~\ref{app:polynomial}.

For $0<\eta<1$ and $C_{ij,\eta}>0$, the physical root is
\begin{equation}
q=C_{ij,\eta}^2.
\end{equation}

It is the smallest of the four roots, since the other three are
\begin{equation}
\begin{split}
&[C_{ij,\eta}+2(r_3+r_4)]^2,\\
&[C_{ij,\eta}+2(r_2+r_4)]^2,\\
&[C_{ij,\eta}+2(r_2+r_3)]^2.
\end{split}
\end{equation}

To establish strictness, note that
\begin{equation}
\begin{aligned}
\rho_{ij,s}
={}&\eta^2\rho_{ij}+(1-\eta)^2\frac{I_4}{4}\\
&+\eta(1-\eta)\left(
\rho_i\otimes\frac{I_2}{2}
+\frac{I_2}{2}\otimes\rho_j
\right)\\
\succeq{}&(1-\eta)^2\frac{I_4}{4}\succ0
\qquad(0<\eta<1).
\end{aligned}
\label{eq:fullrank}
\end{equation}
Here $A\succeq B$ means that $A-B$ is positive semidefinite,
while $A\succ B$ means that $A-B$ is positive definite.
Thus $M_{ij,s}$ is similar to the positive-definite Hermitian matrix
$\rho_{ij,s}^{1/2}\widetilde{\rho}_{ij,s}\rho_{ij,s}^{1/2}$,
so all $r_\alpha$ are strictly positive.
When $C_{ij,\eta}>0$, the other three roots of $\Pi_{ij}$
are strictly larger than $C_{ij,\eta}^2$. Hence
\begin{equation}
\left.
\frac{\partial \Pi_{ij}}{\partial q}
\right|_{q=C_{ij,\eta}^2}
<0.
\label{eq:Piqnegative}
\end{equation}

\section{Invariant deformation and full-trajectory ordering}
\label{sec:ordering}
We now consider how the concurrence polynomial changes along the Kempe fiber.
\begin{lemma}[Invariant deformation]
\label{lem:deformation}
Consider a fixed $\bm E$ and a reduced pair $ij$, and let $k$ denote
the third qubit. Along the corresponding Kempe fiber,
\begin{equation}
\partial_U\Pi_{ij}
=
s^2(1-s)^2
\left(q^2-2L_{ij}q-H_{ij}\right),
\label{eq:PiUgeneral}
\end{equation}
where
\begin{equation}
\begin{split}
L_{ij}
=
\frac14\big[
&1-s(R_i+R_j)\\
&+s^2(1-R_i-R_j-2R_k+4U)
\big],
\end{split}
\label{eq:Lij}
\end{equation}
and
\begin{equation}
H_{ij}=e_1^2-4e_2-8D.
\end{equation}
Here $e_m$ are the elementary symmetric polynomials of the eigenvalues
of $M_{ij,s}$, and $D=\det\rho_{ij,s}$. For any pure three-qubit state
and $0\le s\le1$,
\begin{equation}
L_{ij}\ge0.
\label{eq:Lijpositive}
\end{equation}

At the physical root $q=C_{ij,\eta}^2$, for $0<\eta<1$ and
$C_{ij,\eta}>0$,
\begin{equation}
H_{ij}>q^2,
\end{equation}
so that
\begin{equation}
\left.
\partial_U\Pi_{ij}
\right|_{q=C_{ij,\eta}^2}
<0.
\label{eq:PiUnegativeGeneral}
\end{equation}
\end{lemma}

\begin{proof}
The reduction to Eq.~(\ref{eq:PiUgeneral}) is worked out in
Appendices~\ref{app:reduction} and \ref{app:polynomial}.

Introduce
\begin{equation}
G=1-R_A-R_B-R_C+2U.
\label{eq:Gdef}
\end{equation}
Then
\begin{equation}
4L_{ij}
=
(1-s)^2
+(2-R_i-R_j)s(1-s)
+2Gs^2.
\label{eq:Ldecomposition}
\end{equation}

The first two terms are nonnegative. In the Ac\'{\i}n canonical form
\begin{equation}
|\psi\rangle
=
\lambda_0|000\rangle
+\lambda_1e^{i\phi}|100\rangle
+\lambda_2|101\rangle
+\lambda_3|110\rangle
+\lambda_4|111\rangle,
\label{eq:acinmain}
\end{equation}
where $\lambda_j\geq0$, $\sum_{j=0}^4\lambda_j^2=1$, and
$0\leq\phi\leq\pi$, we have
\begin{equation}
G
=
4\lambda_0^2
\left[
\lambda_4^2
+4\lambda_2^2\lambda_3^2
-4\lambda_1\lambda_2\lambda_3\lambda_4\cos\phi
\right].
\label{eq:Gacin}
\end{equation}
Since $\lambda_1\cos\phi\le1$,
\begin{equation}
G
\ge
4\lambda_0^2
(\lambda_4-2\lambda_2\lambda_3)^2
\ge0.
\label{eq:Gpositive}
\end{equation}
Hence $L_{ij}\ge0$.

The constant term $\Pi_{ij}(0)=H_{ij}^2$ determines only
the magnitude of $H_{ij}$. Its sign follows from the identity
\begin{equation}
H_{ij}
=
-\prod_{\alpha=1}^{4}
\left(
2r_\alpha-\sum_{\beta=1}^{4}r_\beta
\right).
\label{eq:Hsigned}
\end{equation}
When $C_{ij,\eta}>0$, the $\alpha=1$ factor equals
$C_{ij,\eta}$, while the other three factors are negative.
Hence

\begin{equation}
\begin{aligned}
H_{ij}
={}&
C_{ij,\eta}
[C_{ij,\eta}+2(r_3+r_4)]
[C_{ij,\eta}+2(r_2+r_4)]
\\
&\times
[C_{ij,\eta}+2(r_2+r_3)].
\end{aligned}
\label{eq:HfactorGeneral}
\end{equation}
For $0<\eta<1$ and $C_{ij,\eta}>0$, each of the last three factors is
larger than $C_{ij,\eta}$, and therefore
\begin{equation}
H_{ij}>C_{ij,\eta}^4=q^2.
\end{equation}
Together with $L_{ij}\ge0$, Eq.~(\ref{eq:PiUgeneral}) then gives
Eq.~(\ref{eq:PiUnegativeGeneral}).
\end{proof}

The sign above determines the variation of the physical concurrence root
along the Kempe fiber and fixes the ordering given below.

\begin{theorem}[Full-trajectory ordering]
\label{thm:main}
Let $|\psi_1\rangle$ and $|\psi_2\rangle$ be pure three-qubit states
with the same initial entanglement coordinates $\bm E$ and Kempe
invariants $\kappa_1<\kappa_2$.
Under identical, independent local depolarizing channels,
\begin{equation}
C_{ij,\eta}(\kappa_1)
\ge
C_{ij,\eta}(\kappa_2),
\qquad
ij=AB,AC,BC,
\label{eq:mainordering}
\end{equation}
for all $0\le\eta\le1$. For $0<\eta<1$, $C_{ij,\eta}$ decreases
strictly with $\kappa$ as long as the pair remains entangled. Equality
is recovered at $\eta=0,1$, and the non-strict ordering remains valid
once the concurrence has vanished.
\end{theorem}

\begin{proof}
Fix a pair $ij$ with $C_{ij,\eta}>0$ and set
\begin{equation}
q=C_{ij,\eta}^2.
\end{equation}
Since $\Pi_{ij}(q,U)=0$, implicit differentiation gives
\begin{equation}
\frac{\partial q}{\partial U}
=
-\frac{\partial_U\Pi_{ij}}
{\partial_q\Pi_{ij}}.
\end{equation}
At the physical root,
\begin{equation}
\partial_U\Pi_{ij}<0,
\qquad
\partial_q\Pi_{ij}<0,
\end{equation}
by Eqs.~(\ref{eq:PiUnegativeGeneral}) and
(\ref{eq:Piqnegative}). Hence
\begin{equation}
\frac{\partial q}{\partial U}<0.
\end{equation}
Using
\begin{equation}
U=\frac{4\kappa-1}{3},
\end{equation}
we obtain
\begin{equation}
\partial_\kappa C_{ij,\eta}<0
\end{equation}
whenever the pair remains entangled. The same reasoning applies to
$AB$, $AC$, and $BC$.

For fixed $0<\eta<1$, the concurrence is a continuous, single-valued
function of $\kappa$ and decreases strictly wherever it is positive.
Once it reaches zero, it cannot become positive again as $\kappa$
increases. Any subsequent interval of positive concurrence would
have a zero value at its left endpoint, contradicting continuity
and strict decrease within that interval.
The concurrence is therefore nonincreasing over the entire allowed
Kempe interval.

At $\eta=1$, the pairwise concurrences are fixed by $\bm E$ and are
the same for all states on the fiber. At $\eta=0$, every reduced pair
is maximally mixed. Moreover, for $0<\eta_1<\eta_2$,
\begin{equation}
\Lambda_{\eta_1}
=
\Lambda_{\eta_1/\eta_2}
\circ
\Lambda_{\eta_2},
\label{eq:composition}
\end{equation}
so further local depolarization cannot restore pairwise entanglement
after it has disappeared. This completes the ordering for the full
range $0\le\eta\le1$.
\end{proof}

\section{Dynamical consequences}
\label{sec:consequences}
Equation~\eqref{eq:mainordering} implies the same ordering for any
function that is nondecreasing in each of the three pairwise
concurrences. One example is the sum of squared pairwise
entanglements of formation,
\begin{equation}
\mathcal M_{\rm EOF}(\eta)
=
\sum_{i<j} E_f^2(C_{ij,\eta}).
\label{eq:MEOFgeneral}
\end{equation}
Here $E_f(C)$ is the two-qubit entanglement of formation
\cite{Wootters1998},
\begin{equation}
E_f(C)
=
h_2\!\left(\frac{1+\sqrt{1-C^2}}{2}\right),
\label{eq:EOFdefinition}
\end{equation}
with $h_2(x)=-x\log_2 x-(1-x)\log_2(1-x)$ the binary entropy.
Since $E_f(C)$ is nonnegative and increasing,
$\mathcal M_{\rm EOF}$ is nonincreasing in $\kappa$ at fixed
$\bm E$ and $\eta$. It vanishes when all three pairs are separable.

Writing $p=1-\eta$, define $p_{\rm ESD}^{ij}(\kappa)$ as the first value of
$p$ for which the concurrence vanishes, with $p_{\rm ESD}^{ij}=0$ if the
pair is initially separable. Theorem~\ref{thm:main} gives
\begin{equation}
p_{\rm ESD}^{ij}(\kappa_1)\geq p_{\rm ESD}^{ij}(\kappa_2),
\qquad \kappa_1<\kappa_2.
\label{eq:ESDorder}
\end{equation}
The inequality is strict for initially entangled pairs.

To see this, Eqs.~\eqref{eq:de2appendix} and
\eqref{eq:dDappendix} imply
\begin{equation}
\partial_U H_{ij}
=
-\frac12 s^2(1-s)^2<0
\qquad(0<s<1).
\label{eq:HU}
\end{equation}
For a full-rank state, Eq.~\eqref{eq:Hsigned} shows that
$H_{ij}$ has the same sign as $r_1-r_2-r_3-r_4$:
the other three factors are strictly negative.
Thus $H_{ij}>0$ characterizes the entangled region and
$H_{ij}=0$ its boundary.

By continuity, the positive initial concurrence remains positive
for sufficiently small $p$. Since $H_{ij}=-1/16$ at $s=0$
($p=1$), $H_{ij}$ remains negative and the concurrence vanishes
for $p$ sufficiently close to $1$. The first death threshold
therefore lies strictly between $0$ and $1$.

Let $p_2=p_{\rm ESD}^{ij}(\kappa_2)$ and $s_2=(1-p_2)^2$,
and write $U_\ell=(4\kappa_\ell-1)/3$.
Since $H_{ij}(s_2,U_2)=0$, Eq.~\eqref{eq:HU} gives
\begin{equation}
\begin{aligned}
H_{ij}(s_2,U_1)
&=\frac12 s_2^2(1-s_2)^2(U_2-U_1)\\
&>0.
\end{aligned}
\end{equation}
The reduced pair with $\kappa_1$ is therefore entangled at $p_2$
and, by continuity, for slightly larger $p$.
Equation~\eqref{eq:composition} ensures that increasing $p$
amounts to applying further local noise, so an already separable
pair cannot become entangled again. Hence
\begin{equation}
p_{\rm ESD}^{ij}(\kappa_1)>p_{\rm ESD}^{ij}(\kappa_2)
\quad\text{if }C_{ij}>0.
\label{eq:ESDstrict}
\end{equation}

The allowed Kempe interval is
\begin{equation}
\kappa_-(\bm E)\leq\kappa\leq\kappa_+(\bm E).
\label{eq:kappa_interval}
\end{equation}
For physically admissible $\bm E$, set
\begin{equation}
\begin{gathered}
x=C_{AB}^2,\quad y=C_{AC}^2,\quad z=C_{BC}^2,\\
S=x+y+z,\quad P=\sqrt{xyz},\\
Q_0=\sqrt{(t+x)(t+y)(t+z)}.
\end{gathered}
\end{equation}
The static compatibility conditions give \cite{Song2026Compatibility}
\begin{align}
\kappa_-
&=1-\frac34(S+t)+\frac34\max\{-P,Q_0-t\},\\
\kappa_+
&=1-\frac34(S+t)+\frac34P.
\label{eq:KempeEndpoints}
\end{align}
Combining these endpoints with Theorem~\ref{thm:main} gives
\begin{equation}
C_{ij,\eta}[\kappa_+(\bm E)]
\leq C_{ij,\eta}(\kappa)
\leq C_{ij,\eta}[\kappa_-(\bm E)]
\label{eq:generalband}
\end{equation}
for every $\eta$. The lower Kempe endpoint simultaneously maximizes
the concurrence of all three pairs, while the upper endpoint
minimizes it.

As an example, take
\begin{equation}
\bm E=(0.55,0.45,0.35,0.25).
\label{eq:numericalE}
\end{equation}
Table~\ref{tab:fiberexample} gives the noisy concurrences at the lower
endpoint, midpoint, and upper endpoint of the Kempe interval.
At fixed noise strength, all three concurrences decrease as
$\kappa$ increases. The $AB$ sudden-death thresholds are approximately
$0.293449$ at $\kappa_-$ and $0.290013$ at $\kappa_+$.

At fixed $\bm E$ and noise strength $0<\eta<1$, the concurrence
of a chosen pair is strictly decreasing in $\kappa$ wherever
it is positive. A nonzero value therefore uniquely determines
$\kappa$, while leaving the discrete complex-conjugation ambiguity
unresolved. Different Kempe values that yield zero concurrence
cannot be distinguished by this quantity alone.

\begin{table}[t]
\caption{Concurrences at $\eta=0.9$ for the fixed initial coordinates in
Eq.~\eqref{eq:numericalE}. Rows correspond to the lower Kempe endpoint,
the midpoint, and the upper endpoint. All values are rounded to six
decimal places.}
\label{tab:fiberexample}
\begin{ruledtabular}
\begin{tabular}{cccc}
$\kappa$ & $C_{AB,0.9}$ & $C_{AC,0.9}$ & $C_{BC,0.9}$\\
\hline
0.383251 & 0.326159 & 0.241538 & 0.160299\\
0.395047 & 0.323727 & 0.238959 & 0.157503\\
0.406844 & 0.321328 & 0.236415 & 0.154752
\end{tabular}
\end{ruledtabular}
\end{table}

\section{Discussion}
\label{sec:discussion}

Along a fixed-$\bm E$ Kempe fiber, the initial local Bloch-vector
lengths and $K_{ij}=\|T^{ij}\|_F^2$ are fixed. The reduced-state
purities are also independent of $\kappa$ at each noise strength.
To examine the correlations that can still vary, we write the
initial pair-correlation tensor in principal-axis form,
\begin{equation}
T^{ij}
=
O_i\,\mathrm{diag}(t_1,t_2,t_3)\,O_j^T.
\end{equation}
Here $O_i,O_j\in SO(3)$, and the $t_a$ are signed singular values,
with $K_{ij}=\sum_a t_a^2$. In the corresponding local frames,
the Bloch vectors are
\begin{equation}
\bm u=O_i^T\bm r_i,
\qquad
\bm v=O_j^T\bm r_j.
\end{equation}
Thus
\begin{equation}
U
=
\bm r_i^T T^{ij}\bm r_j
=
\sum_{a=1}^{3} t_a u_a v_a.
\label{eq:alignmentU}
\end{equation}
Appendix~\ref{app:reduction} gives
\begin{equation}
\begin{aligned}
J_{ij}
&=\|\operatorname{cof}T^{ij}\|_F^2
=\sum_{a<b}t_a^2t_b^2,\\
\partial_U J_{ij}
&=-2,\qquad \det T^{ij}=U-R_k.
\end{aligned}
\label{eq:spectralchange}
\end{equation}
The nonzero derivative of $J_{ij}$ shows that the singular-value
spectrum changes along a nondegenerate fiber.
Thus the variation in Eq.~\eqref{eq:alignmentU} cannot be attributed
solely to changes in the relative orientations of the local Bloch
vectors and the principal axes.

Local depolarization is LU covariant and rescales the local Bloch
vectors and pair-correlation tensors as
$\bm r_i\mapsto\eta\bm r_i$ and
$T^{ij}\mapsto\eta^2T^{ij}$, respectively.
These different scalings change the relative weights of the
one- and two-body terms in the reduced state.
The common ordering of all three pairs is established by the
polynomial argument in Eqs.~\eqref{eq:PiUgeneral}
and \eqref{eq:Piqnegative}.
The spectral and moment identities used in this argument rely
on the initial three-qubit state being pure.

The ordering also holds at each time for any common time dependence
$\eta=\eta(T)\in[0,1]$, even when $\eta(T)$ is nonmonotonic.
Equation~\eqref{eq:composition} rules out concurrence revival
while $\eta(T)$ decreases, whereas revival is possible when
$\eta(T)$ increases. Here $T$ denotes time, while $t=\tau_3$ remains the fixed initial
three-tangle.

\section{Conclusion}
\label{sec:conclusion}

For pure three-qubit states with the same initial pairwise
concurrences and three-tangle, the Kempe invariant orders the
concurrence trajectories of all three pairs under identical,
independent local depolarizing channels.
The smaller the Kempe invariant, the more resistant pairwise
entanglement is to depolarizing noise, and the stronger the noise
required to destroy it completely. The lower and upper endpoints
of the allowed Kempe interval therefore bound all three concurrence
trajectories from above and below, respectively.

The next question is whether the ordering survives when the three qubits experience 
different depolarizing strengths, described by $\Lambda_{\eta_A}\otimes\Lambda_{\eta_B}\otimes\Lambda_{\eta_C}$ 
with $0\leq\eta_A,\eta_B,\eta_C\leq1$. For other noise models, such as anisotropic 
Pauli noise, amplitude damping, and thermal channels, one needs to determine when the 
concurrence trajectories remain ordered and when they cross. Such comparisons must also 
specify the local orientations of the initial states if the channels are not LU covariant. 
Extending the result to mixed initial states is another open problem.
Beyond concurrence, a separate question is whether the Kempe invariant
orders Bell nonlocality or quantum steering, since their ordering
cannot in general be inferred from that of concurrence.

\acknowledgments
This work was supported by the Natural Science Foundation of Anhui Province under Grant No. 2508085ZD001, and Scientific Research Projects of Anhui Provincial Department of Education under Grant Nos. 2025AHGXZK20126 and 2025AHGXZK50054.
\appendix

\section{Invariant reduction for a general reduced pair}
\label{app:reduction}

We again take $AB$ as the representative pair and write
\begin{equation}
T\equiv T^{AB},
\qquad
U=\bm r_A^TT\bm r_B,
\qquad
K=\|T\|_F^2.
\end{equation}
For the representative pair $AB$, Eq.~(\ref{eq:Kij}) reads
$K=1+2R_C-R_A-R_B$.

We use the identity
\begin{equation}
\det T=U-R_C.
\label{eq:dgeneral}
\end{equation}

This result also follows directly from the Bloch representation. For a
general two-qubit state of the form Eq.~(\ref{eq:blochAB}), we have
\begin{equation}
\operatorname{Tr}\rho_{AB}^3
=
\frac1{16}
\left[
1+3(R_A+R_B)+3K+6U-6\det T
\right].
\label{eq:trrho3pair}
\end{equation}

For the pure state $|\psi\rangle_{ABC}$, $\rho_{AB}$ and $\rho_C$ have
the same nonzero eigenvalues, so that
\begin{equation}
\operatorname{Tr}\rho_{AB}^3
=
\operatorname{Tr}\rho_C^3
=
\frac{1+3R_C}{4}.
\end{equation}

Substituting this relation into the previous expression gives
Eq.~(\ref{eq:dgeneral}).

For the channel calculation it is convenient to introduce
\begin{align}
J&=\|\operatorname{cof}T\|_F^2,
\\
W_A&=\bm r_A^TTT^T\bm r_A,
\\
W_B&=\bm r_B^TT^TT\bm r_B,
\\
V&=\bm r_A^T(\operatorname{cof}T)\bm r_B.
\label{eq:JWVVdefs}
\end{align}
Here $\operatorname{cof}T$ denotes the cofactor matrix. We only need the
dependence of these quantities on the fiber coordinate $U$. The remaining
terms can be expressed in terms of the fixed coordinates $\bm E$.
Evaluating the invariants in the Ac\'{\i}n form of
Eq.~(\ref{eq:acinmain}), with
\begin{equation}
a=C_{AB},\qquad b=C_{AC},\qquad c=C_{BC},
\end{equation}
one obtains the following quantities, which are constant along the
fixed-$\bm E$ fiber:
\begin{align}
J+2U
={}&
a^4-a^2b^2-a^2c^2+b^2c^2
\nonumber\\
&-2b^2-2c^2-2t+2,
\label{eq:Jcombo}
\\
W_A-2U
={}&
a^2+b^4+b^2t+c^2+t-1,
\label{eq:WAcombo}
\\
W_B-2U
={}&
a^2+b^2+c^4+c^2t+t-1,
\label{eq:WBcombo}
\\
V-U
={}&
a^4+a^2t+b^2+c^2+t-1.
\label{eq:Vcombo}
\end{align}
The right-hand sides depend only on the fixed entanglement coordinates,
so that
\begin{equation}
\begin{aligned}
\partial_U(\det T)&=1,
&\partial_UJ&=-2,\\
\partial_UW_A=\partial_UW_B&=2,
&\partial_UV&=1.
\end{aligned}
\label{eq:staticderivatives}
\end{equation}
The formulas for the other two pairs follow by cyclic permutation.

\section{Auxiliary polynomial and Kempe deformation}
\label{app:polynomial}

Let $e_1,e_2,e_3,e_4$ denote the elementary symmetric polynomials of the
eigenvalues $\mu_\alpha=r_\alpha^2$ of $M_{ij,s}$, and define
\begin{equation}
D=\det\rho_{ij,s},
\qquad
e_4=D^2.
\end{equation}
Define
\begin{equation}
q_\alpha
=
\left(2r_\alpha-\sum_\beta r_\beta\right)^2.
\end{equation}
Using the symmetric polynomials, one finds
\begin{align}
\sum_\alpha q_\alpha
&=4e_1,
\\
\sum_{\alpha<\beta}q_\alpha q_\beta
&=6e_1^2-8e_2+48D,
\\
\sum_{\alpha<\beta<\gamma}q_\alpha q_\beta q_\gamma
&=64e_3+4e_1(e_1^2-4e_2+8D),
\\
\prod_\alpha q_\alpha
&=(e_1^2-4e_2-8D)^2.
\end{align}
Hence
\begin{align}
\Pi_{ij}(q)
={}&
q^4-4e_1q^3
+\left(6e_1^2-8e_2+48D\right)q^2
\nonumber\\
&-64e_3q
-4e_1\left(e_1^2-4e_2+8D\right)q
\nonumber\\
&+\left(e_1^2-4e_2-8D\right)^2 .
\label{eq:Piexplicit}
\end{align}

We next derive the $U$ derivatives used in
Lemma~\ref{lem:deformation}.
We again use the pair $AB$ and set
\begin{equation}
\chi=1-s(R_A+R_B)+s^2K.
\end{equation}
The spin-flipped product can be written as
\begin{equation}
M_s
=
\frac1{16}
\left[
\chi I+\sum_{m,n}Q_{mn}\sigma_m\otimes\sigma_n
\right],
\label{eq:M_Qform}
\end{equation}
where
\begin{equation}
\begin{split}
Q=2s\big[&T-\bm r_A\bm r_B^T-s\operatorname{cof}T\\
&+i\sqrt{s}([\bm r_A]_\times T+T[\bm r_B]_\times^T)\big].
\end{split}
\label{eq:Qmatrix}
\end{equation}
Here $[\bm r]_\times$ denotes the antisymmetric cross-product matrix.
Along the fiber, $R_A$, $R_B$, and $K$ are fixed, so
\begin{equation}
\partial_Ue_1=0,
\qquad
e_1=\frac{\chi}{4}.
\end{equation}

Writing
\begin{equation}
Q:Q\equiv\sum_{mn}Q_{mn}Q_{mn},
\end{equation}
where the colon denotes algebraic contraction without complex
conjugation, direct contraction yields
\begin{equation}
Q:Q=4s^2\Xi,
\end{equation}
with
\begin{equation}
\begin{split}
\Xi={}&K+R_AR_B-2U+s^2J\\
&+s\big[W_A+W_B-(R_A+R_B)K\\
&\hspace{1.2cm}-6\det T+6V\big].
\end{split}
\end{equation}
Using Eq.~(\ref{eq:staticderivatives}),
\begin{equation}
\partial_U(Q:Q)=-8s^2(1-s)^2.
\end{equation}
Since
\begin{equation}
e_2=\frac{3\chi^2-Q:Q}{128},
\end{equation}
we obtain
\begin{equation}
\partial_Ue_2=\frac{s^2(1-s)^2}{16}.
\label{eq:de2appendix}
\end{equation}

Using the moments
$p_n=\operatorname{Tr}\rho_{AB,s}^n$,
the determinant $D$ can be written in terms of
\begin{align}
p_2
&=\frac14[1+s(R_A+R_B)+s^2K],
\\
p_3
&=\frac1{16}\big[1+3s(R_A+R_B)+3s^2K
\nonumber\\
&\hspace{1.5cm}+6s^2U-6s^3\det T\big],
\\
p_4
&=\frac1{64}\Big\{[1+s(R_A+R_B)+s^2K]^2
\nonumber\\
&\quad+4s(R_A+R_B)+4s^2(K+R_AR_B+6U)
\nonumber\\
&\quad+4s^3(W_A+W_B-6\det T-2V)+4s^4J\Big\}.
\label{eq:p4}
\end{align}

Equation~(\ref{eq:staticderivatives}) then gives
\begin{align}
\partial_Up_2&=0,
\\
\partial_Up_3&=\frac38s^2(1-s),
\\
\partial_Up_4&=\frac18s^2(1-s)(3+s).
\end{align}

For a unit-trace $4\times4$ matrix,
\begin{equation}
D
=
\frac{1-6p_2+3p_2^2+8p_3-6p_4}{24},
\end{equation}
and therefore
\begin{equation}
\partial_UD=\frac{s^2(1-s)^2}{32}.
\label{eq:dDappendix}
\end{equation}

Finally, Pauli algebra gives
\begin{equation}
\operatorname{Tr}
\left(\sum_{mn}Q_{mn}\sigma_m\otimes\sigma_n\right)^3
=-24\det Q,
\end{equation}
and hence
\begin{equation}
e_3
=
\frac{\chi^3-\chi(Q:Q)-2\det Q}{1024}.
\label{eq:e3Q}
\end{equation}

To make the determinant reduction explicit, use
$a=C_{AB}$, $b=C_{AC}$, and $c=C_{BC}$ as in
Appendix~\ref{app:reduction}, and define
\begin{equation}
\begin{aligned}
f_0&=a^2(a^2+t),\\
f_1&=1-b^2-c^2-2a^2-2t,\\
f_2&=-1+2b^2+2c^2-4a^2,\qquad d=\det T.
\end{aligned}
\label{eq:detQcoefficients}
\end{equation}
Expansion of Eq.~(\ref{eq:Qmatrix}) in the Ac\'{\i}n form, followed by
the invariant substitutions in Appendix~\ref{app:reduction}, yields
\begin{equation}
\begin{split}
\det Q=8s^3\big[&f_0(-1+2f_1s+f_2s^2)\\
&-s(1-s)^2d^2\big].
\end{split}
\label{eq:detQexpanded}
\end{equation}
All three $f_m$ are fixed by $\bm E$, while $\partial_Ud=1$.
Differentiating at fixed $s$ gives

\begin{equation}
\partial_U\det Q
=-16s^4(1-s)^2\det T.
\label{eq:detQderivative}
\end{equation}

Combining this result with Eq.~(\ref{eq:dgeneral}), we have
\begin{equation}
\partial_Ue_3
=
\frac{s^2(1-s)^2}{32}L_{AB},
\label{eq:de3appendix}
\end{equation}
where
\begin{equation}
L_{AB}
=
\frac14[1-s(R_A+R_B)+s^2(1-R_A-R_B-2R_C+4U)].
\end{equation}
The other two pairs follow by cyclic permutation.
Substituting
Eqs.~(\ref{eq:de2appendix}), (\ref{eq:dDappendix}), and
(\ref{eq:de3appendix}) into Eq.~(\ref{eq:Piexplicit}) gives
Eq.~(\ref{eq:PiUgeneral}).

\section{Boundary and degenerate cases}
\label{app:boundaries}

The invariant expressions in Appendices~\ref{app:reduction}
and \ref{app:polynomial} contain no division by state parameters
that may vanish and are continuous in the state amplitudes.
The corresponding identities therefore extend to nongeneric
pure states by taking limits of generic states, even where
the Ac\'{\i}n parametrization is degenerate.

For $0<\eta<1$, Eq.~\eqref{eq:fullrank} guarantees that every
reduced pair is full rank. Whenever $C_{ij,\eta}>0$,
Eq.~\eqref{eq:Piqnegative} ensures that
$q=C_{ij,\eta}^2$ is a simple root of $\Pi_{ij}$.
The implicit-differentiation argument therefore remains valid
at interior points of a nondegenerate Kempe interval, without
requiring differentiable canonical coordinates or individually
differentiable spin-flip eigenvalues.

Continuity extends the non-strict ordering to the Kempe endpoints.
If the interval collapses to a single point, there are no distinct
Kempe values to compare. The noise endpoints and the
zero-concurrence region are treated in the proof of
Theorem~\ref{thm:main}.


\end{document}